\documentclass[preprint,12pt]{elsarticle}

\usepackage{graphicx}

\usepackage{hyperref}

\usepackage{amssymb}
\usepackage{amsmath}

\journal{Nuclear Physics A}

\begin{document}

\begin{frontmatter}



\title{Pion, muon decays and weak interaction symmetries}


\author{D. Po\v{c}ani\'c\corref{cor1}}
\ead{pocanic@virginia.edu}
\cortext[cor1]{Corresponding author}
\author{(for the PIBETA Collaboration)\corref{cor2}}
\address{Department of Physics, University of Virginia,
  Charlottesville, VA 22904-4714, USA}

\begin{abstract}
We review the recent measurements of the rare pion decays:
$\pi^+\to\pi^0e^+\nu$ (beta, $\pi_{e3}$, or $\pi_\beta$ decay),
radiative decay $\pi^+\to e^+\nu\gamma$ ($\pi_{e2\gamma}$ or RPD), and
$\pi^+\to e^+\nu$ ($\pi_{e2}$), as well as the radiative muon decay,
$\mu \to e\nu\bar{\nu}\gamma$, their theoretical implications, and
prospects for further improvement.
\end{abstract}


\end{frontmatter}




Thanks to exceptionally well controlled theoretical uncertainties,
decays of light mesons, particularly of the pion, are understood at
very high precision, typically a few parts per $10^4$, or better (see,
e.g., Refs.~\cite{Jau01,Mar06,Cir07}.  Hence, pion decays present
fertile ground for testing predictions of the standard model (SM), as
well as for setting constraints on processes and particles outside the
SM.  Muon decays are theoretically cleaner yet, and provide direct
information concerning the symmetry properties of the weak interaction
itself, e.g., departures from its $V$$-$$A$ form.

\begin{figure}[!ht]
  \parbox{0.65\linewidth}{\includegraphics[width=\linewidth]{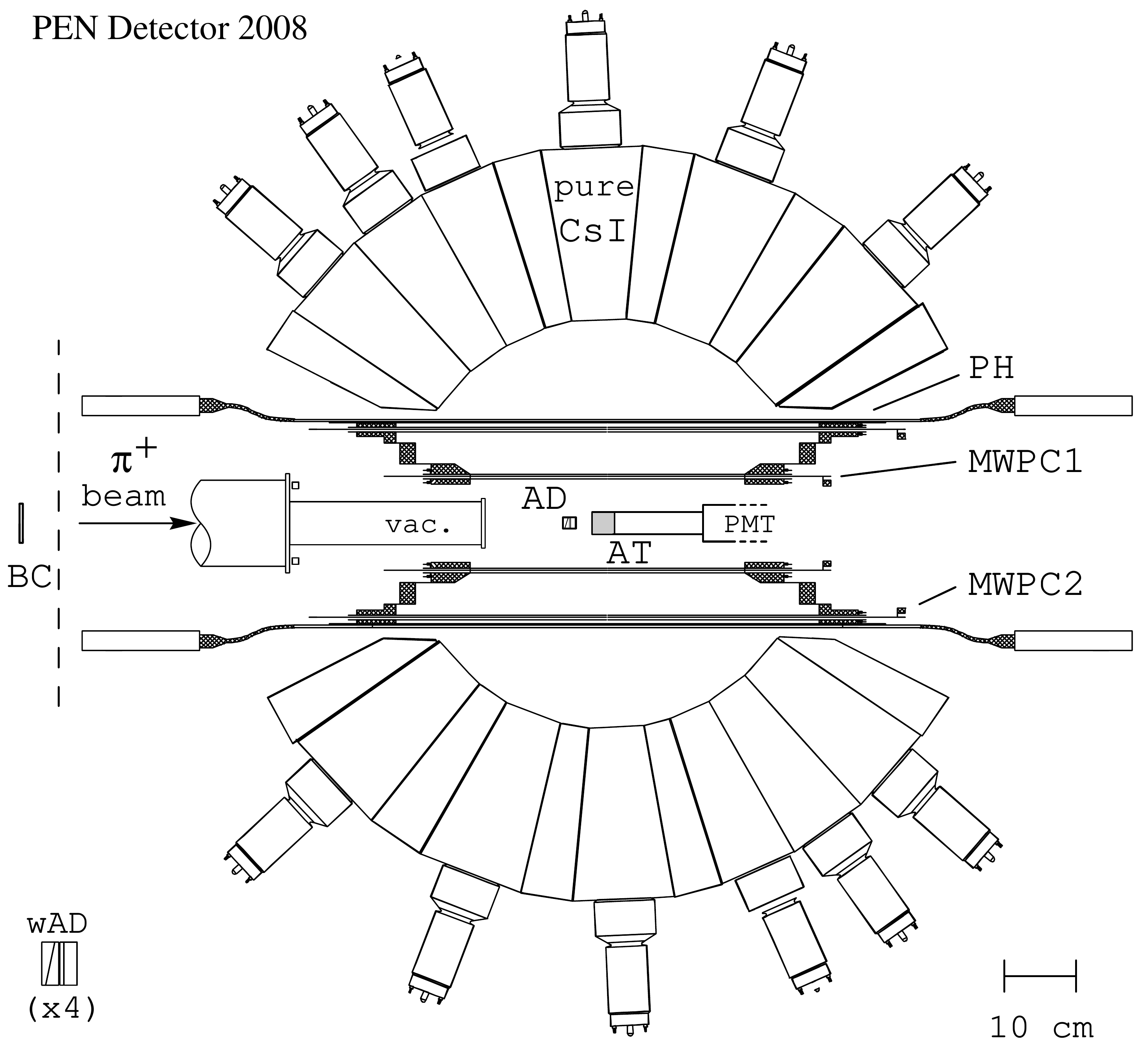}}
  \hspace*{\fill}
  \parbox{0.33\linewidth}{\caption{Schematic cross section of the
      PIBETA/PEN apparatus, shown here in the 2008 PEN run
      configuration, with its main components: beam entry with the
      upstream beam counter (BC), wedged active degrader (wAD) and
      target (AT), cylindrical multiwire proportional chambers
      (MWPC's), plastic hodoscope (PH) detectors and photomultiplier
      tubes (PMT's), 240-element pure CsI electromagnetic shower
      calorimeter and its PMT's.  BC, wAD, and PH detectors are made
      of plastic scintillator.}
      \label{fig1}  }
\end{figure}

The PIBETA experiment, with measurements in 1999--2001 and 2004 at the
Paul Scherrer Institute (PSI), Switzerland, was primarily designed to
improve the accuracy of the $\pi_\beta$ decay branching ratio.  Pion
decays at rest were detected in an detector system \cite{Frl04a,PBweb}
shown schematically in Fig.~\ref{fig1}.
Normalizing to the number of observed $\pi^+\to e^+\nu$ ($\pi_{e2}$)
decays, we determined the branching ratio value $B^{\rm ex-n}
(\pi^+\to\pi^0e^+\nu) = 1.036(4)_{\rm stat}(4)_{\rm syst}(3)_{\rm e2}
\times 10^{-8}$, where the first uncertainty is statistical, the
second systematic, and the third arises from $\Delta B(\pi_{\rm e2})$,
experimental $\pi_{e2}$ branching ratio uncertainty\cite{Poc04} .
Normalizing instead to the more precise theoretical value of
$B(\pi_{\rm e2})$ \cite{Cir07} yields $B^{\rm th-n} (\pi^+\to \pi^0
e^+ \nu) = 1.040(4)_{\rm stat}(4)_{\rm syst }\times 10^{-8}$.  Both
results agree well with the SM prediction, and represent the best test
to date of vector current conservation (CVC) in a meson.

Concurrently with the $\pi_{e3}$ decay, the PIBETA collaboration has
measured the $\pi^+\to{\rm e}^+\nu\gamma$ (RPD) branching ratio over a
wide region of phase space.  Two sets of amplitudes contribute to RPD:
the inner-bremsstrahlung, IB, fully described by QED, and the
structure-dependent amplitude, SD.  The standard V$-$A electroweak
theory requires only two pion form factors, $F_A$, axial vector, and
$F_V$, vector, to describe the SD amplitude.  The vector form factor
is strongly constrained by the CVC hypothesis to $F_V = 0.0259(9)$.

\begin{figure}[!ht]
  \parbox{0.55\linewidth}{\includegraphics[width=\linewidth]{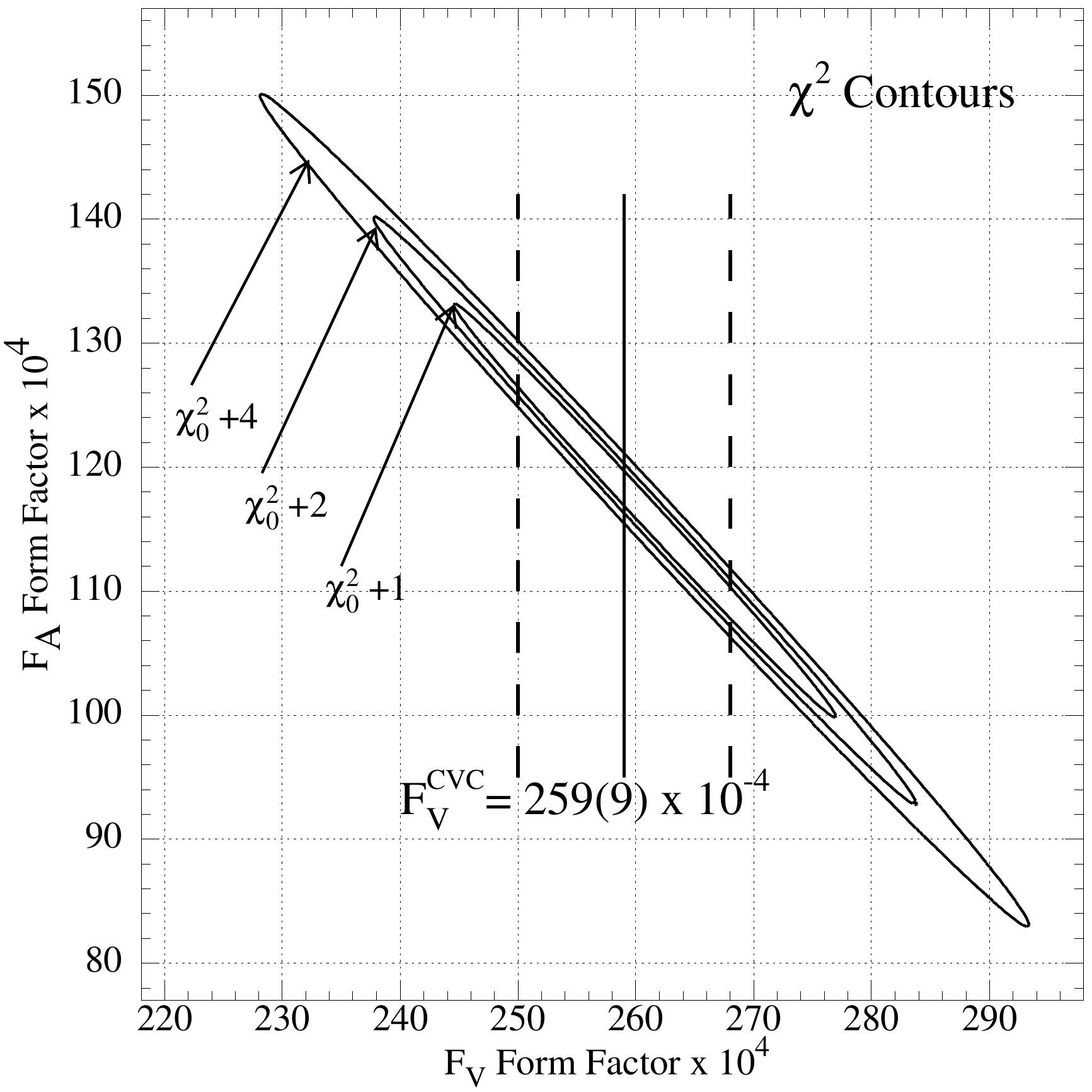}}
  \hspace*{\fill}
  \parbox{0.43\linewidth}{\caption{Contour plot of loci of constant
      $\chi^{2}$ for the minimum value $\chi_0^2$ plus 1, 2, and 4
      units, respectively, in the $F_A$-$F_V$ parameter plane, keeping
      the parameter $a=0.041$.
      The range of the CVC prediction, $F_V = 0.0259(9)$ is indicated.}
      \label{fig2}  }
\end{figure}

Minimum-$\chi^2$ fits to the measured $(E_{e^+},E_\gamma)$ energy
distributions result in the weak form factor value of $F_A=0.0119(1)$
with a fixed value of $F_V=0.0259$.  An unconstrained fit yields
$F_V=0.0258(17)$ and $F_A=0.0117(17)$ in a tightly correlated narrow
band (see Fig.~\ref{fig2}).  
In addition, we have measured $a=0.10(6)$ for the dependence of $F_V$
on $q^2$, the ${\rm e}^{+}\nu$ pair invariant mass squared,
parametrized as $F_V(q^2)=F_V(0)(1+a\cdot q^2)$.  The branching ratio
for the kinematic region $E_\gamma > 10\,$MeV and $\theta_{{\rm e^+}
  \gamma} > 40^\circ $ is measured to be $B^{\rm exp}=73.86(54) \times
10^{-8}$.  Earlier deviations we found in the
high-$E_\gamma$/low-$E_{{\rm e}^+}$ kinematic region \cite{Frl04b} are
resolved, and we find full compatibility with CVC and standard
$V$$-$$A$ calculations without a tensor term---we find $-5.2 \times
10^{-4} < F_T < 4.0 \times 10^{-4}$ with 90\% condidence.  We also
derive new values for the pion polarizability at leading order,
$\alpha_E = \rm 2.78(10) \times 10^{-4}\,fm^3$, and neutral pion
lifetime, $\tau_{\pi 0} = (8.5 \pm 1.1) \times 10^{-17}\,$s.

Radiative muon decay (RMD) offers a particularly sensitive means for
testing the $V$$-$$A$ nature of the weak interaction through
$\bar{\eta}$, the only Michel parameter not accessible in ordinary
$\mu$ decay.  RMD events with energetic photons are required to
evaluate $\bar{\eta}$.  Along with other Michel parameters,
$\bar{\eta}$ sets limits on departures from the $V$$-$$A$ weak
interaction form ($\bar{\eta}_{SM} \equiv 0$).

With more than $4 \times 10^5$ RMD events, the 2004 PIBETA data set
leads to a preliminary result: $ B(E_\gamma > 10\,{\rm MeV},\,
\theta_{e\gamma} > 30^\circ) = 4.40(2)_{\rm stat}(9)_{\rm syst} \times
10^{-3}$, 14 times more precise than the previous world
average\cite{PDG08}; here the systematic uncertainty will be reduced
by improving low-energy photon scattering simulation.  The best fit
for $B$ is obtained for $\bar{\eta} = \rm -0.084 \pm 0.050 (stat.) \pm
0.034 (syst.)$, yielding upper limits on the allowed value of
$\bar{\eta} \leq 0.033$ and $\bar{\eta} \leq 0.060$, with 68\% and
90\% confidence, respectively.  Combined with previous measurements of
$\bar{\eta}$, this reduces the known upper limit by a factor of 2.5 to
$\bar\eta_{\,\text{\sc world avg}} \leq 0.028$, with 68\% confidence
\cite{Van05}.


Historically, the $\pi$$\to$$e\nu$ (or $\pi_{e2}$) decay, provided an
early strong confirmation of the $V$$-$$A$ nature of the electroweak
interaction.  At present, its branching ratio is understood at the
level of better than one part in $10^4$ \cite{Cir07}.  Experimental
precision, however, lags behind by over an order of magnitude.
Because of the large helicity suppression of the $\pi_{e2}$ decay, its
branching ratio is highly susceptible to even slight non-$V$$-$$A$
contributions from new physics, making this decay a particularly
suitable subject of study.

The PEN experiment \cite{PENweb} uses a modified PIBETA detector
system to carry out a measurement of $B(\pi_{e2})$ to an accuracy of
$\Delta B/B \leq 5 \times 10^{-4}$, at PSI.  During engineering runs
in 2007 in 2008 the collaboration developed the required intense low
energy pion beam tunes and upgraded key detector components, including
a mini time projection chamber to map the beam.  To date, the
experiment has observed over $7 \times 10^{10}$ tagged pion stops in
the target, and recorded over $4 \times 10^6$ $\pi_{e2}$ decays before
cuts.  PEN will run in 2009-10 in order to complete the required event
statistics and key systematic studies.

In conclusion, the PIBETA experiment has improved, by an order of
magnitude or better, the accuracy of rare decays: the pion beta
($\pi_{e3}$), radiative pion ($\pi_{e2\gamma}$), and radiative muon
decays.  In doing so, PIBETA has verified CVC and SM predictions at
new levels in a meson, and improved the precision of pion structure
parameters ($F_V$, $F_A$, $F_T$ form factors, polarizability, closely
related to low-energy effective QCD lagrangian parameters
$L_9^r+L_{10}^r$).  Through the $B(\mu\to e\nu\bar{\nu}\gamma)$ and
$\bar{\eta}$ measurement we have made possible new tests of the
$V$$-$$A$ nature of the weak interaction.  PEN, the successor
experiment, is well on its way to improving the precision of the
$\pi_{e2}$ branching ratio, with the prospect of setting new limits on
light lepton universality, and on a variety of non-SM processes
manifested primarily through pseudoscalar contributions.  The PEN data
set will more than double our 2004 PIBETA run statistics for RPD and
RMD events, leading to further improvements in precision, while the
$\pi_\beta$ event statistics will increase only slightly.









\setlength{\parskip}{0pt}
\setlength{\parsep}{0pt}

\end{document}